# Programmable spectral shaping to improve the measurement precision of frequency comb mode-resolved spectral interferometric ranging

Yoon-Soo Jang,[1,2,*], Sunghoon Eom,[1] Jungjae Park,[1,2] and Jonghan Jin,[1,2,3]

[1]Length Standard Group, Division of Physical Metrology, Korea Research Institute of Standards and Science (KRISS), 267 Gajeong-ro, Yuseong-gu, Daejeon, 34113, Republic of Korea

[2]Department of Science of Measurement, University of Science and Technology (UST), Daejeon, 34113, Rep. of Korea

[3]jonghan@kriss.re.kr

*ysj@kirss.ac.kr

**Abstract**

Comb-mode resolved spectral domain interferometry (CORE-SDI), which is capable of measuring length of kilometers or more with precision on the order of nanometers, is considered to be a promising technology for next-generation length standards, replacing laser displacement interferometers. In this study, we aim to improve the measurement precision of CORE-SDI using programmable spectral shaping. We report the generation of effectively broad and symmetric light sources through the programmable spectral shaping. The light source used here was generated by the spectrally-broadened electro-optic comb with a repetition rate of 17.5 GHz. Through the programmable spectral shaping, the optical spectrum was flattened within 1 dB, resulting in a square-shaped optical spectrum. As a result, the 3-dB spectral width was extended from 1.15 THz to 6.7 THz. We performed a comparison between the measurement results of various spectrum shapes. We confirmed an improvement in the measurement precision from 69 nm to 6 nm, which was also corroborated by numerical simulations. We believe that this study on enhancing the measurement precision of CORE-SDI through the proposed spectral shaping will make a significant contribution to reducing the measurement uncertainty of future CORE-SDI systems, thereby advancing the development of next-generation length standards.

## 1. Introduction

Length, as one of the fundamental physical quantities, plays a crucial role in range of different areas, including science [1] and technology [2]. The current standard length unit, the meter, is defined as follows: "The metre, symbol $m$, is defined by taking the fixed numerical value of the speed of light in vacuum $c$ to be 299 792 458 when expressed in the unit m s$^{-1}$" [3]. The *mises en pratique* regarding the definition of the meter are implemented using laser displacement interferometers based on laser wavelengths [4,5]. However, due to the $2\pi$ ambiguity problem, it is necessary to accumulate displacements from the initial position to the target position in order to measure absolute lengths, which imposes several practical limitations on the use of this approach [6]. To overcome these limitations, various absolute distance measurement methodologies have been developed to enable the determination of distances in a single measurement. These include multi-wavelength interferometry [7,8], wavelength scanning interferometry [9,10], frequency-modulated continuous wave (FMCW) ranging [11,12], amplitude-modulated continuous wave (AMCW) ranging [13,14], and others. However, these approaches typically have a measurement precision of around 1 μm, which is significantly lower compared to the nanometer-level precision achieved by laser displacement interferometers [15]. Spectral domain interferometry (SDI) is capable of measuring absolute distances with nanometer-level precision, indicating its potential to replace laser displacement interferometers. However, conventional SDI commonly relies on a broadband spectrum source, such as a super-luminescent diode, which imposes limitations on the coherence length. As a result, the practical application of conventional SDI is essentially limited to fields such as optical coherence tomography (OCT) and thickness measurements, which utilize the millimeter-scale measurement range [16,17].

Recently, the advent of frequency combs has led to significant advancements in absolute distance measurement research trends [18]. While frequency combs possess a broadband spectrum in the frequency domain, individual modes of the frequency comb exhibit very narrow linewidths. The frequency combs can be synchronized with frequency standards, enabling them to achieve extremely high accuracy. In the time domain, they exhibit narrow pulse widths at the femtosecond level. This characteristic has not only revolutionized distance measurements but has also led to significant advancements in optical metrology [19,20]. The use of frequency combs for absolute distance measurements has been advanced through various methodologies, including multi-wavelength interferometry [21-23], AMCW ranging [24-26], FMCW ranging [27,28], dual-comb ranging [29-31], time-of-flight measurements [32,33], and SDI [34-38]. Particularly from the perspective of SDI, frequency comb-based SDI can maintain nanometer-level precision even for long-distance measurements due to the ability of this technology to provide a broadband spectrum and a long coherence length [39-44].

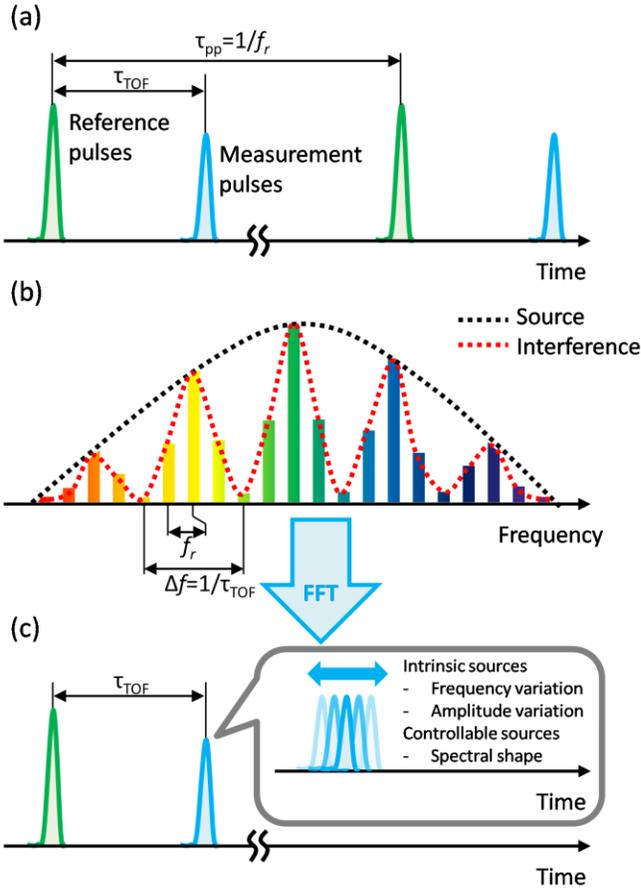

**FIG. 1. Conceptual image describing the basic principle of a frequency comb based spectral domain interferometer.** (a) Reference and measurement pulses of the frequency comb separated by $\tau_{TOF}$ in the time domain. (b) Spectral interference that occurs when the reference and measurement pulses of the frequency comb are combined in the frequency domain. (c) Reconstructed time domain signal obtained through Fourier transformation and some of the causes of the temporal fluctuation observed in the signal.

Regarding the principle of SDI (see Appendix A for details), the time delay ($\tau_{TOF}$) caused by the optical path difference is represented in the frequency domain as the period of the interference signal (1/ $\tau_{TOF}$) [45]. The interference signal in the frequency domain can be Fourier transformed to measure the time delay in the time or length domain. In frequency comb-based SDI, factors that can influence the measurement precision include intrinsic sources such as frequency noise and amplitude noise. Frequency noise can be considered negligible in frequency comb-based SDI given that frequency combs typically have frequency uncertainty on the order of $10^{-12}$ or better [46]. Previous research has identified amplitude noise as an intrinsic factor limiting precision of frequency comb-based SDI [37,38]. One controllable factor is the spectral shape of the light source. A broader spectrum of the light source leads to a narrower peak signal when using the Fourier transform, enabling a higher temporal resolution to be achieved [38,47]. If the spectral shape is asymmetric, the peak signal will also exhibit asymmetry, potentially distorting the peak position. Therefore, a symmetric shape is preferred to minimize distortions in the peak position [48].

With regard to OCT, several studies attempted to adjust the spectral shape to achieve high-resolution images [49-52]. However, there is currently a lack of research exploring the impact of these adjustments on the precision of absolute distance measurement. In this study, we report on the impact of programmable spectral shaping on the measurement precision of comb mode resolved spectral domain interferometry (CORE-SDI). By employing programmable spectral shaping to modify the spectrum of the frequency comb into various shapes, we aim to enhance the measurement precision. We also provide a detailed analysis of the resulting changes. We utilized highly nonlinear fiber (HNLF) to achieve spectral broadening and employed an electro-optic comb (EO comb) with a repetition rate of 17.5 GHz, which allows for individual comb modes to be spectrally resolved. Through the use of a waveshaper and data post-processing, we successfully generated 382 comb modes in a square waveform with 1-dB flatness, facilitating easy application of various windows. As a result, we achieved an expansion of the 3-dB bandwidth of the spectral width from 1.15 THz to 6.7 THz. We applied various window functions, in this case Gaussian, $Sech^2$, and Supergaussian, with varying spectral widths to the experimental data to analyze their effect on the measurement precision. We confirmed that the measurement precision improved from 69 nm to 6 nm after programmable spectral shaping. Furthermore, these findings were consistently observed through numerical simulations. We have positive expectations that the method proposed here can make a significant contribution to the replacement of laser displacement interferometry with frequency comb-based spectral interferometry and advance the progress in next-generation length standard.

## 2. Results and discussion
### 2.1. Optical layout of frequency comb based spectral interferometry

Figure 2 shows the optical configuration of frequency comb based spectral interferometry with programmable spectral shaping. A supercontinuum electro-optic comb (SC EO comb) is generated using a typical configuration with a repetition rate of 17.5 GHz [53,54]. The SC EO comb consists of more than 400 frequency modes, with the majority of the spectral density concentrated around a center wavelength of 192 THz. As a result, the typical spectral bandwidth is 1.15 THz for a 3 dB cutoff and 4 THz for a 10 dB cutoff. The left section in fig. 2(b) shows the optical spectrum of the supercontinuum EO comb. When using an optical spectrum analyzer (OSA) with high sensitivity but a slow measurement speed, it is possible to observe more than 400 frequency modes of the SC EO comb. However, when using a spectrometer with fast measurement capabilities but a limited dynamic range, the device tends selectively to observe signals with high amplitudes near the center frequency. Consequently, only a subset of the multiple frequency modes can be observed. In this study, a spectrometer was used for high-speed measurements. To enable the measurement of a large number of frequency modes with the spectrometer used here, a wave shaper was utilized to transform the signal into a square waveform. This effectively expanded the 3-dB spectral bandwidth to 6.7 THz (382 comb modes) [55].

The interferometer used here is composed of fiber-optic components. Reference pulses were generated using 4% of the Fresnel light reflected from the end face of an FC/PC fiber ferrule. The measurement pulses were obtained by reflecting the light back from a measurement mirror positioned at approximately 100 mm. The spectral interference signal was acquired using a high-speed spectrometer (S-Nova-1550, METERLAB) capable of resolving the individual frequency components of the SC EO comb at an update rate of 40 kHz. The measured interference signal was programmatically processed using software to perform spectral shaping, with the goal of improving the CORE-SDI measurement precision.

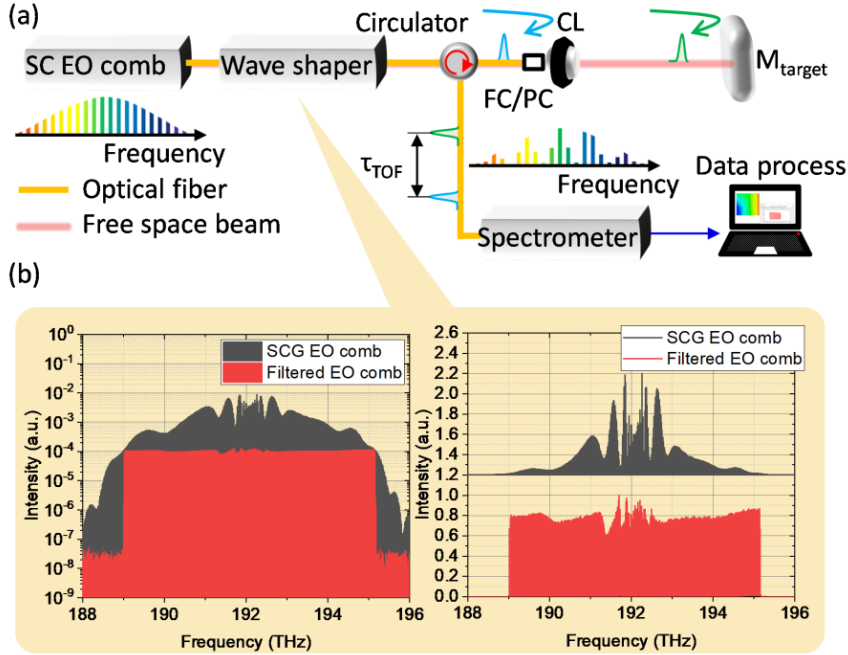

**FIG. 2. Measurement setup of CORE-SDI with programmable spectral shaping.** (a) Optical layout of the proposed method. FC/PC: Face contact with physical contact. CL: collimating lens. M: mirror. (b) Optical spectrum before and after the wave shaper on the log scale (left) and linear scale (right).

### 2.2. Programmable spectral shaping of the EO comb

Figure 3 depicts the process of programmable spectral shaping through post-data processing. The spectral interference signal generated by the SC EO comb (green line) was confined within a limited spectral range, whereas the spectral interference signal generated by the filtered comb (gray line) using the wave shaper spanned a wide spectral bandwidth. However, the spectral flatness of the filtered spectrum was not sufficiently uniform due to hardware limitations. To achieve programmable spectral shaping using various window functions, a square waveform spectrum is preferred. To do this, the spectrum of the filtered comb (red line) was measured, and the interference spectrum was subsequently software-normalized to generate a square waveform interference signal (blue line) with flatness within 1 dB. The signals outside of the square waveform were digitally set to zero. As a result, spectral bandwidth increased from 1.15 THz to 6.7 THz.

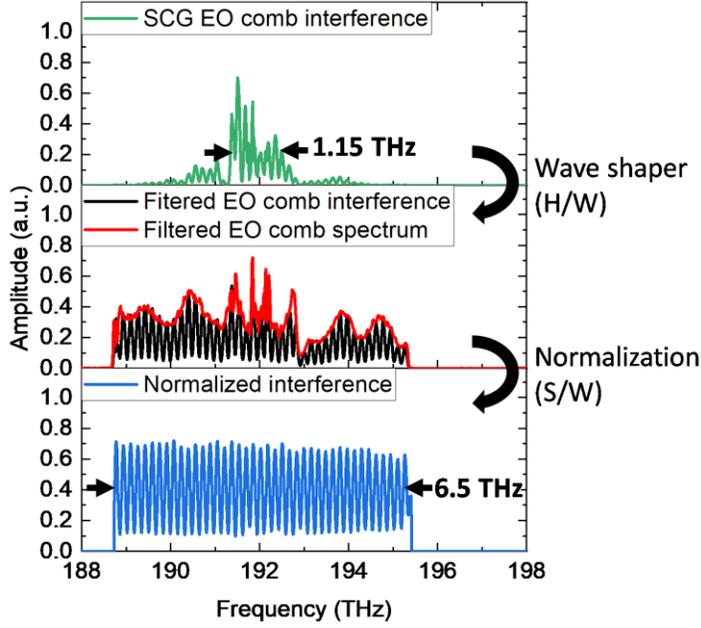

**FIG. 3. Spectral flattening by a wave shaper and post-data processing.** Hardware and software based spectral flattening will cause the spectral bandwidth to increase from 1.15 THz to 6.7 THz.

Figure 4(a) depicts the spectral interference signal before and after the application of programmable spectral shaping. We applied window functions, in this case the Gaussian and Sech$^2$ shapes, which represent the typical spectrum of an ideal pulsed laser, as well as a 10th-order Supergaussian window function with a flat-top-like shape. The 3-dB spectral bandwidth of the window functions was applied up to 6 THz. In the figure below, for convenience, we show the 2, 4, and 6 THz cases. A limited number of frequency comb modes within the 6.7 THz range can restrict the visibility of frequency components beyond a certain spectral bandwidth. For the Gaussian and Sech$^2$ window functions, when the 3-dB spectral bandwidth exceeds 3 THz, the frequency components outside that range may not be observable or may have reduced visibility.

The result of applying Fourier transformation to the programmable spectral-shaped interference and scaling it up in the time domain to reconstruct the length-domain signal is shown in Fig. 4(b). Before programmable spectral shaping, the length-domain signal showed distortions including side peaks, background noise, and other artifacts in addition to the main peak components. For the Gaussian and Sech$^2$ window functions with a 3-dB spectral bandwidth of 2 THz, clean peak signals were observed at the peak positions without any pedestals. However, for the 3-dB spectral bandwidths of 4 THz and 6 THz, pedestals appeared on the sides of the peak signals. This phenomenon occurred because the frequency components beyond the observed spectral range were not captured, resulting in a distorted representation of the frequency spectrum. The Supergaussian window function offers the advantage of generating an interference spectrum that closely matches the shape of the window function. However, due to its similarity to a square waveform, the peak signal in the length domain exhibited multiple pedestals. In all window functions, it was observed that a wider spectral width resulted in a narrower peak component in the length domain.

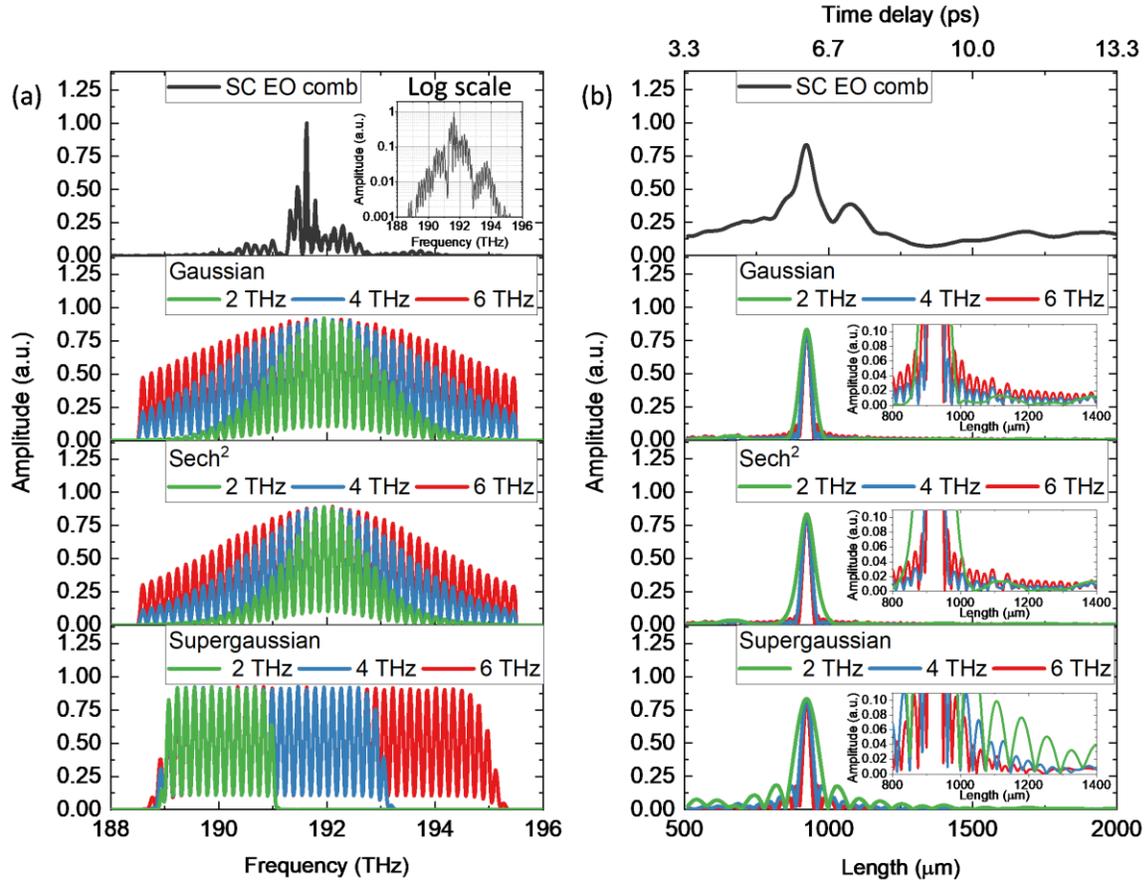

**FIG. 4. Programmable spectral shaping and corresponding Fourier transformed signal in the length domain.** (a) Spectral interference before and after programmable spectral shaping. Gaussian, Sech$^2$ and 10th-order Supergaussian window functions were applied with 3-dB spectral bandwidth from 2 to 6 THz. (b) Reconstructed length-domain signal before and after programmable spectral shaping. Insets show their magnified graph.

## 2.3. Numerical simulation of the measurement precision of frequency comb based spectral interferometry

The peak position in the length domain is determined through polynomial curve fitting, as described in Appendix B. The intensity fluctuation of the spectral interference influences the amplitude variation of the Fourier transformed signal. The intensity fluctuation of the spectral interference can indeed affect the determination of the peak position through polynomial curve fitting. As shown in Fig. 5, the effects of the spectral bandwidth and intensity fluctuation on the measurement precision were investigated through numerical simulations, which were conducted considering Gaussian, $Sech^2$, and Supergaussian window functions, with fluctuations ranging from 0.0001 (0.01%) to 0.1 (10%) and 3-dB spectral bandwidths ranging from 1 THz to 6 THz as shown in Fig. 5 (a) to (c). The simulated precisions were determined by standard deviation value using 1,000 data points. In all three window function case, i.e., Gaussian, $Sech^2$, and Supergaussian, it was observed that decreasing the intensity fluctuation and widening the spectral bandwidth led to an improvement in the measurement precision. The predicted impact of intensity fluctuations on the measurement precision was found to be similar for all three cases. It was observed that when the intensity fluctuation was decreased by a factor of 10, the measurement precision improved by a factor of 10. Fig. 5(d) presents the simulation results of measurement precision with respect to intensity fluctuations for the spectral bandwidth of 6 THz as an example.

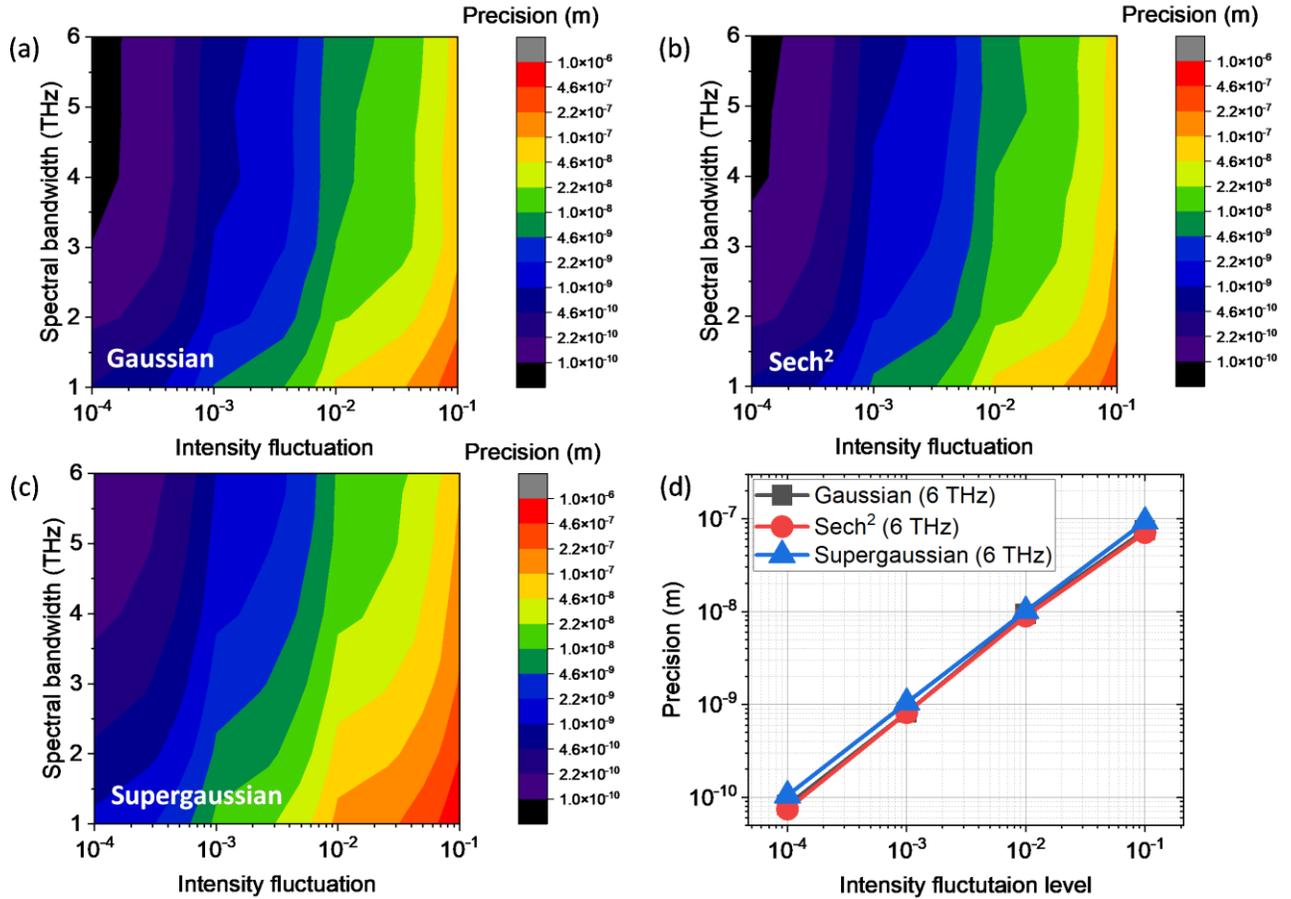

**Fig 5. Numerical estimation of measurement precision by a numerical simulation.** (a), (b) and (c) Estimated measurement precision with intensity fluctuations and spectral bandwidths for Gaussian, Sech$^2$ and Supergaussian shapes, respectively. (d) Estimated measurement precision with intensity fluctuations for Gaussian, Sech$^2$ and Supergaussian shapes with a 6 THz spectral bandwidth.

### 2.4. Improved measurement precision by programmable spectral shaping

To investigate the impact of the spectral shape and width on the measurement precision, experimental analysis was conducted by positioning a measurement mirror at a distance of approximately 100 mm as shown in Fig. 6(a). The experimental measurement precision was determined by the standard deviation value using 1,000 data points. Before programmable spectral shaping (SC EO comb shape), the measurement precision was observed to be 68.74 nm. The measurement precision was obtained by applying Gaussian (red line), Sech$^2$ (blue line), and Supergaussian (green line) window functions with spectral bandwidths ranging from 1 THz to 6

THz through programmable spectral shaping, as shown in Figure 6(a). For all window functions (Gaussian, Sech$^2$ and Supergaussian shape), the measurement precision improves with an increase in the spectral bandwidth. With the same spectral bandwidth, the measurement precision was slightly higher when using the Supergaussian window function compared to the Gaussian or Sech$^2$ window functions. This likely occurred because the Supergaussian window function exhibits a lower number of frequency modes compared to the Gaussian or Sech$^2$ window functions, despite having the same spectral bandwidth. For the Supergaussian window function, the measurement precision improves consistently as the spectral bandwidth increases. However, for the Gaussian and Sech$^2$ window functions, the measurement precision did not improve beyond approximately 7 nm for spectral bandwidths exceeding 3 THz. This phenomenon arises due to the absence of information about frequency components outside that range, considering the use of 382 frequency comb modes in this study. To overcome this limitation, utilizing a greater number of frequency modes would be necessary, although doing so may introduce trade-offs in terms of system complexity. In the experimental conditions of this study, the best measurement precision of 5.75 nm was achieved with the Supergaussian window function with a spectral bandwidth of 6 THz.

The experimental results were then compared with the numerical simulation results. The intensity noise was measured and found to be 0.6% RMS, and this value was taken into account in the numerical simulations. As shown in Fig. 6(b), the predicted measurement precision from the numerical simulation exhibits a trend similar to the actual measurement precision obtained from the experiments.

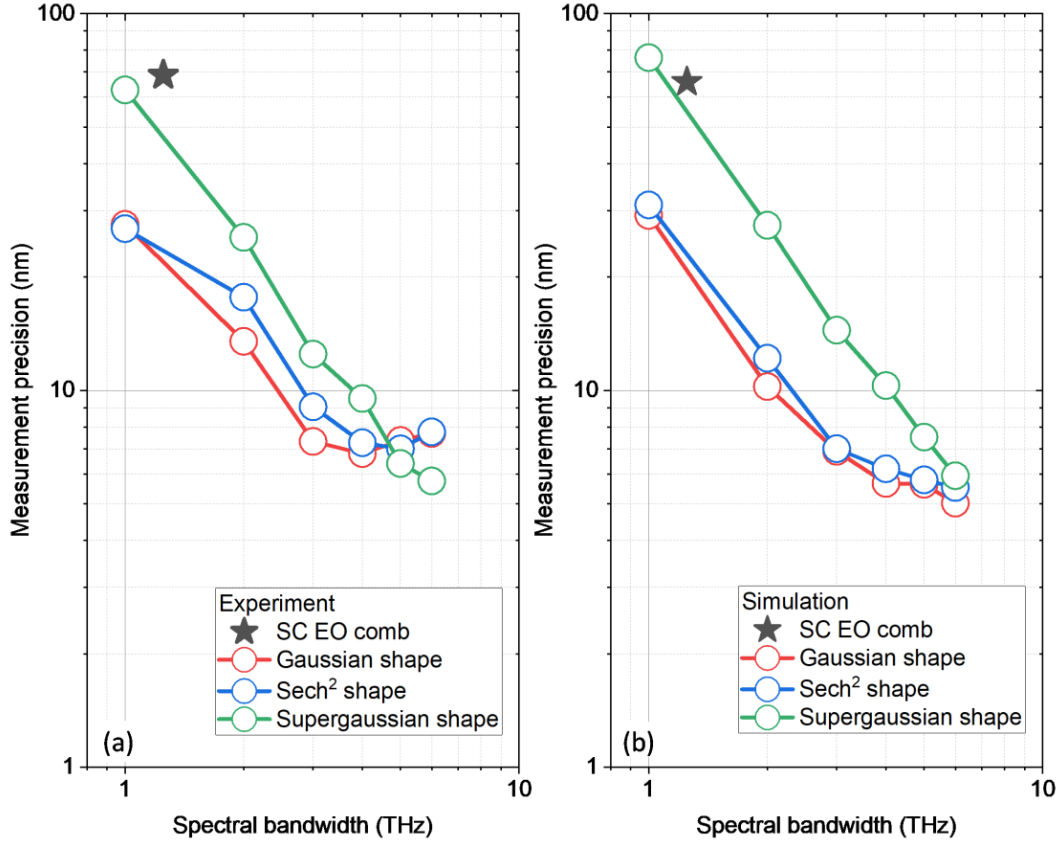

**FIG. 6. Comparison between experimental and simulation results in terms of the measurement precision.** (a) Experimental analysis of the measurement precision, and (b) numerical simulation of the measurement precision.

To demonstrate the potential effects of improved measurement precision through programmable spectral shaping, fixed lengths were measured with an update rate of 40 kHz over 600 ms, resulting in a total of 24,000 data points. Before the application of programmable spectral shaping, the measurement results exhibited significant fluctuations, as indicated by the gray line in Figure 7(a). However, after the application of spectral shaping, the fluctuations in the measurement results were significantly reduced, as indicated by the red line. This improvement in the measurement precision can also be observed in the histogram, where the distribution becomes narrower and more concentrated. Additionally, the target mirror was attached to a PZT stage to introduce vibrations with a frequency of 2 kHz and an amplitude of 30 nm as shown in Fig. 7(b). The length was then measured under this vibrating condition. For comparison and validation purposes, a homodyne laser interferometer (blue line) was also used to measure the vibrations. Before programmable spectral shaping, as shown in Fig. 7(b) (gray line) the vibrational motion was not visible due to the large fluctuations in the measured distance. However, after applying programmable spectral shaping with the Supergaussian window function, the vibrational motion (red line) was clearly observed and showed good agreement with the measurements from the homodyne laser interferometer.

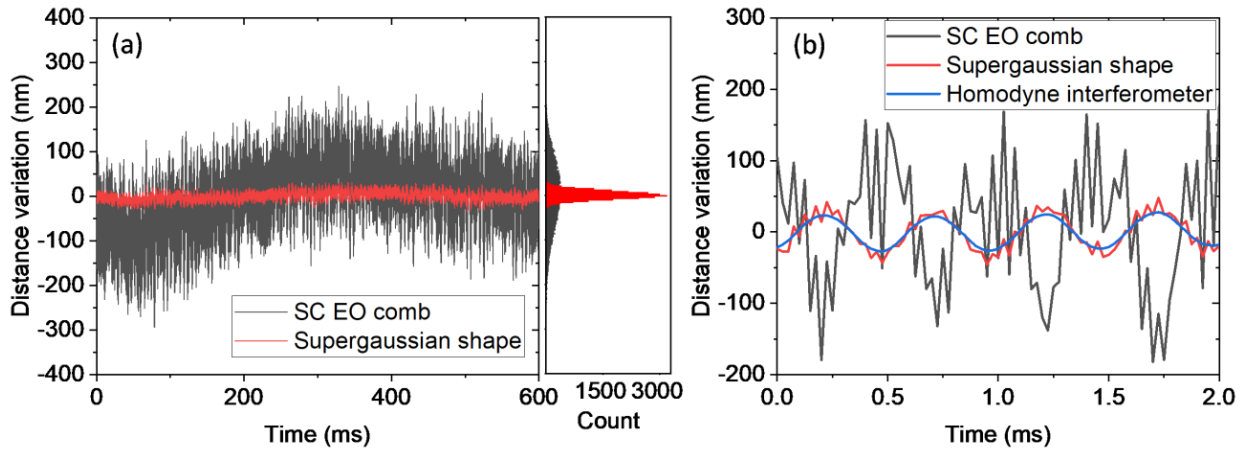

**FIG. 7. Improved measurement precision by programmable spectral shaping.** (a) Time-dependent variation of the measured distance over 600 ms and corresponding histograms. (b) Time-dependent variation of the measured distance under vibration with frequency of 2 kHz and amplitude of 30 nm.

3. **Conclusion**

We proposed programmable spectral shaping to improve the measurement precision of frequency comb mode resolved spectral domain interferometry. Spectral filtering performed in hardware and spectral normalization performed in software expanded the 3-dB spectral bandwidth from 1.15 THz to 6.7 THz. The measurement precisions depending on various spectral shapes were experimentally obtained and were also numerically simulated. Through programmable spectral shaping, the measurement precision was improved from 69 nm to 6 nm, and this phenomenon was observed in the simulations as well.

In principle, an ideal Gaussian or $Sech^2$ spectrum would be the most desirable for spectral shaping. However, in order to achieve an ideal spectrum, an infinitely wide spectral bandwidth of the frequency comb would be required. Due to such a practical constraint, the Supergaussian shaped spectrum, which maximizes the utilization of the spectrum within a limited number of comb modes, is also considered suitable for achieving high-precision distance measurements. It is expected that the programmable spectral shaping method proposed in this study will greatly improve the measurement precision of frequency comb based spectral interferometry, leading to a significant enhancement in measurement uncertainty.

## 4. Appendix A: Principle of spectral domain interferometry

Figure 1 shows basic principle of frequency comb based spectral domain interferometry. [45] In the frequency domain, a frequency comb consists of multiple evenly spaced wavelengths distinguished by the repetition rate ($f_r$). In the time domain, the frequency comb is represented as a series of evenly spaced pulses with a temporal interval between consecutive pulses ($\tau_{pp} = 1/f_r$), where $c$ is the speed of light. In general, the time delay ($\tau_{TOF}$) for the round-trip distance caused by the light reflected from an object located at position $L$ can be expressed as $\tau_{TOF} = 2n_{air}L/c$, where $n_{air}$ is the refractive index of air. In general, when directly measuring $\tau_{TOF}$, the precision is limited to a few picoseconds due to electronics-related constraints. This corresponds to a measurement precision of a few millimeters in length [56]. In spectral domain interferometry, $\tau_{TOF}$ is indirectly measured by utilizing the signals obtained in the spectral domain to measure it more precisely. As shown in Fig. 1(a) and (b), when two pulses (reference pulse and measurement pulse) separated by $\tau_{TOF}$ interact, they produce spectral interference with a periodicity of $1/\tau_{TOF}$. The spectral interference can be measured using commercially available spectrometer. The measured spectral interference is utilized to reconstruct the time-domain signal through Fourier transformation. $\tau_{TOF}$ is determined by the peak position of the reconstructed time-domain signal.

## 5. Appendix B: Data processing for precise peak detection

For a precise determination of the peak position $\tau_{TOF}$ in the Fourier domain, several methods have been proposed, including polynomial fitting [57], the centroid method [58], and the phase-slope method [59]. These methods provide more precise peak detection compared to simply selecting the amplitude peak of the Fourier transform, which is limited by the resolution of the transform. In our case, we utilized the polynomial fitting method due to its rapid processing time, as shown in Fig.8. The peak signal in the Fourier domain can be modeled as $I(\tau) = A\tau^2 + B\tau + C$. In our case, typically 10 points around the peak position were used. The peak position is determined by finding the point where the first derivative is zero, which is given by $dI(\tau)/d\tau = 2A\tau + B = 0$. Thus, the peak position can be determined simply as $\tau = -B/2A$.

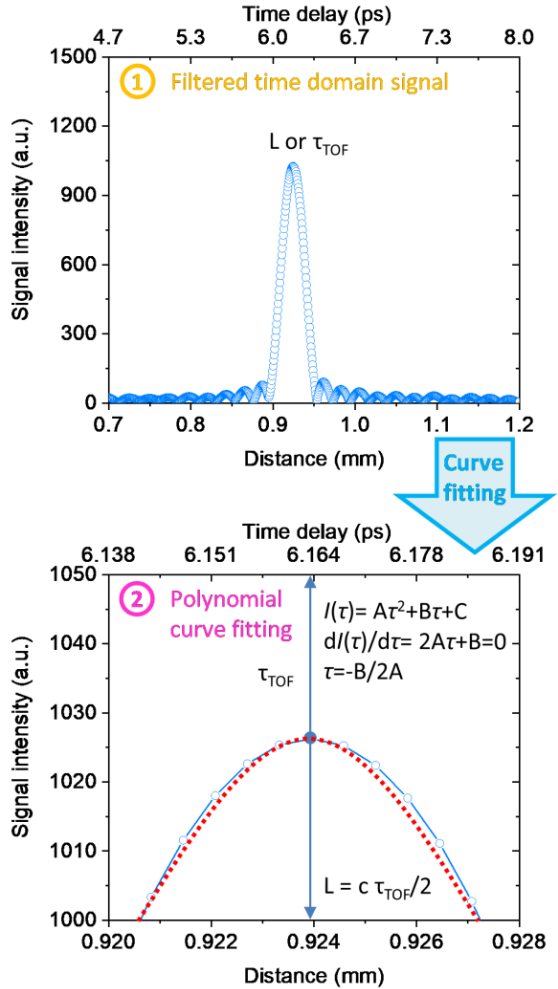

**Fig. 8. Data processing of precise peak detection by polynomial curve fitting**

## 6. Appendix C: Broadband EO comb generation

The EO comb was generated using a typical method involving a CW laser as a seed laser, an intensity modulator, and three phase modulators in a series arrangement [53]. All modulators were driven by the same RF synthesizer, which was synchronized with a GPS-based frequency reference. The typical method of EO comb generation typically provides a spectral bandwidth of only a few nanometers. To overcome this limitation, the EO comb was amplified to 2 W and then passed through a 100 m high-nonlinear fiber for supercontinuum generation (SC) [54]. As shown in Fig. 9(c), the SC EO comb was spectrally expanded from 1530 nm to 1590 nm.

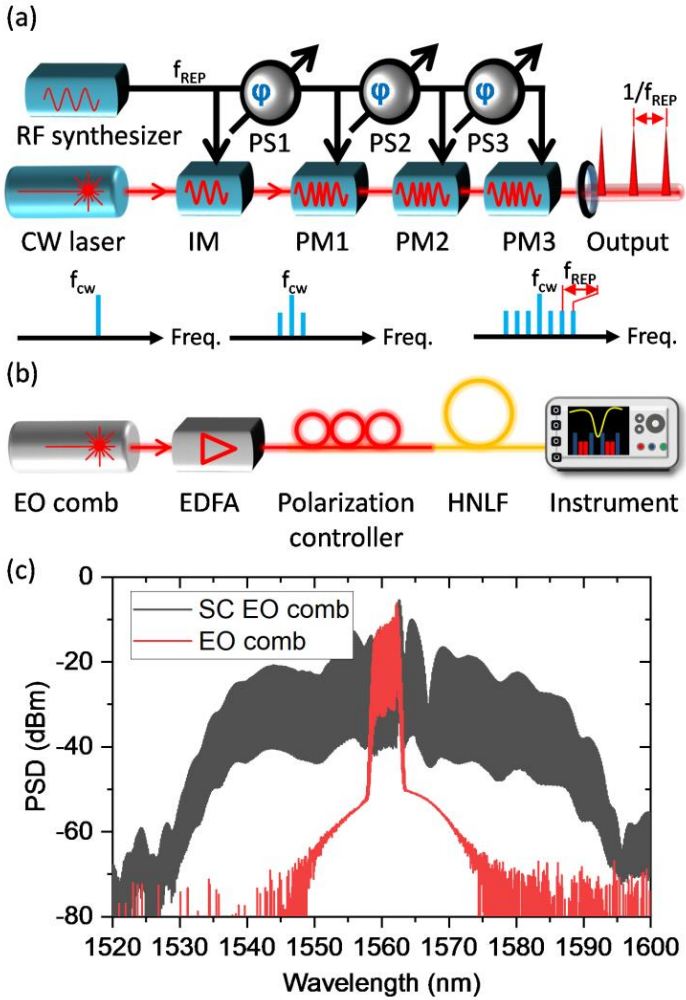

**Fig. 9. Broadband EO comb generation.** (a) EO comb generation by the typical method. (b) Spectral broadening by supercontinuum generation. (c) Optical spectrum before and after supercontinuum generation.

## 7. Appendix D: Experimental results of distance measurements depending on thespectral shape and spectral bandwidth

Figure 10 shows several examples of experimental results for distance measurements using various spectral shapes (Gaussian, Sech$^2$, and Supergaussian) and spectral bandwidths (2 THz, 4 THz and 6 THz) for 25 ms (1,000 data points).

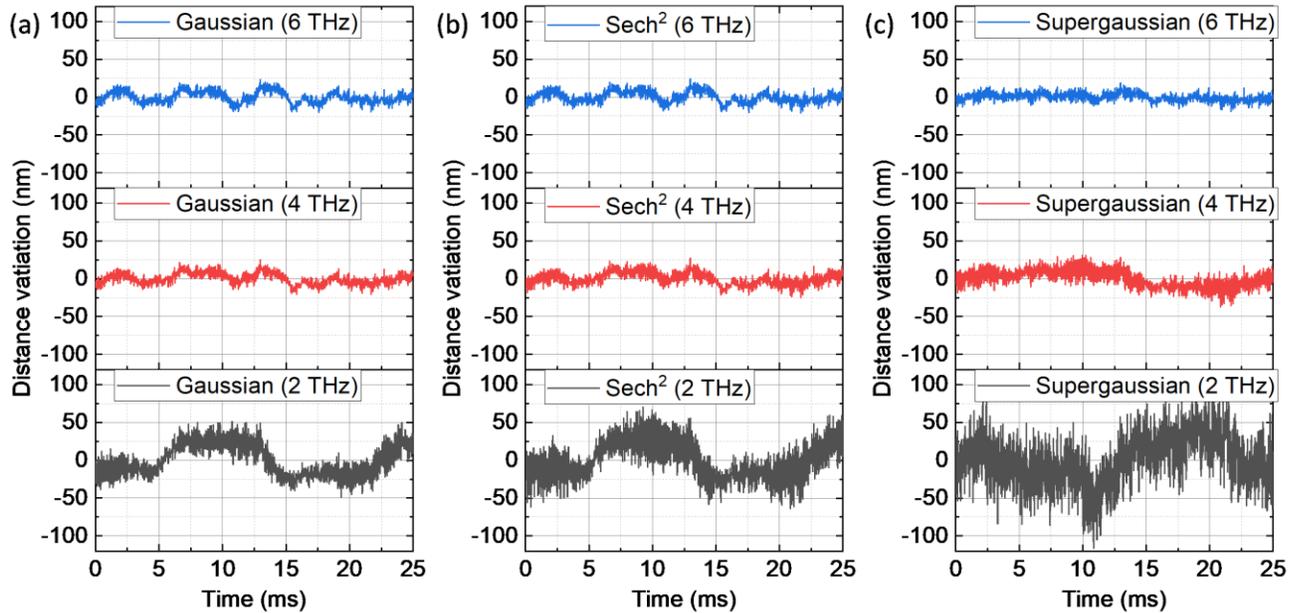

**Fig. 10. Experimental results of distance measurements depending on the spectral shape and spectral bandwidth.** (a) Gaussian window function. (b) Sech$^2$ window function. (c) Supergaussian window function.

**Acknowledgments**: The authors thanks Meterlab Corp. for supporting data acquisition of the high-speed spectrometer. This work is supported by Korea Research Institute of Standard and Science (23011361, 23011041).

**Author contributions:** Y.-S. J. and J. J. led the project. Y.-S. J. designed the experiment. Y.-S. J. performed the electro-optic frequency comb generation and distance measurements. S. Eom performed spectral flattening of the EO comb. Y.-S. J. performed programmable spectral shaping of the EO comb. Y.-S. J., and J. P. conducted the data process of distance measurement. Y.-S. J., J. P. and J. J. conducted the analysis of the measured data. All authors prepared the manuscript.

**Data availability**: The data that support the plots within this paper and other finding of this study are available from the corresponding author upon reasonable request.